\documentclass[pra,twocolumn,superscriptaddress,showpacs,floatfix,nofootinbib]{revtex4}
\usepackage{epsfig,amsmath,amsfonts,amssymb}
\usepackage{graphicx}
\newcommand{\beq}{\begin{equation}}
\newcommand{\eeq}{\end{equation}}

\newcommand{\beqa}{\begin{eqnarray}}
\newcommand{\eeqa}{\end{eqnarray}}
\def\ra{\rangle}
\def\la{\langle}
\begin{document}
\title{Classical picture of post-exponential decay}

\author{E. Torrontegui}
\affiliation{Departamento de Qu\'{\i}mica F\'{\i}sica, Universidad 
del Pa\'{\i}s Vasco - Euskal Herriko Unibertsitatea, Apdo. 644, 48080 
Bilbao, Spain} 

\author{J. G. Muga}
\affiliation{Departamento de Qu\'{\i}mica F\'{\i}sica, Universidad 
del Pa\'{\i}s Vasco - Euskal Herriko Unibertsitatea, Apdo. 644, 48080 
Bilbao, Spain} 

\author{J. Martorell}
\affiliation{Departament d'Estructura i Constituents 
de la Materia, Univ. Barcelona, 08028 Barcelona, Spain}

\author{D. W. L. Sprung}
\affiliation{Department of Physics and Astronomy, 
McMaster University, Hamilton, Ontario L8S 4M1 Canada}
\date{\today}


\begin{abstract}
Post-exponential decay of the probability density of a quantum particle 
leaving a 
trap can be reproduced accurately, 
except for interference oscillations at the transition to the 
post-exponential regime, by means of an ensemble of classical 
particles emitted with constant probability per unit time 
and the same half-life as the quantum system. 
The energy distribution of the ensemble is chosen to be identical to 
the quantum distribution, 
and the classical point source is located at the scattering length of the 
corresponding quantum system. A 1D example is provided to illustrate the 
general argument.

\end{abstract}
\maketitle
\section{Introduction}
Exponential decay is ubiquitous in quantum physics and constitutes
the typical 
dynamical pattern for unstable systems. Theory predicts deviations at 
short times, related to the Zeno effect, and also at long times, with 
significant implications for cosmology \cite{KD08}, hidden variables \cite{Winterh}, 
non-hermitian formulations \cite{NB77,Nicolaides02}, and 
radioactive-dating methods \cite{No95}. On the experimental side, the 
post-exponential regime, generally algebraic, has been elusive. There 
are few claims to have observed it \cite{RHM06}, which has 
engendered much effort to explain why, or to improve its observability 
\cite{book,TMMS}. 

Classically, exponential decay arises from a constant decay probability per unit time. 
What has been lacking so far has been a correspondingly simple, physically appealing 
classical picture of post-exponential decay. The purpose of this short article is to 
provide such a framework, by following up on an old suggestion of R. G. Newton \cite{Newton61,Newtonbook}. 

The standard quantum mechanical derivations of post-exponential decay are not very helpful 
in suggesting such a picture. The early reliance upon the Paley-Wiener theorem \cite{Khalfin}, 
is not very illuminating from a physical perspective and it provides bounds, not precise predictions. 
Similarly post-exponential decay can be attributed mathematically to the fact that the pole contribution 
to exponential decay is eventually comparable to or smaller than a line integral, whose value arises 
predominantly from a saddle point at threshold, associated with slow particles \cite{Petzold,Winter,book}. 
This is surely more intuitive, yet not fully satisfying for those seeking a  pictorial, 
rather than a complex variable, understanding of the phenomenon \cite{Lan}. 
In this vein, Hellund proposed an electrostatic analog 
\cite{Hellund53} which relates quantum emission of radiation to the 
damped oscillation of a charge describable in purely classical 
(stochastic) terms, and interprets the deviation from 
exponential decay as a ``straggling phenomenon'' characteristic of 
a diffusion process. However, quantum dynamics cannot generally be 
reduced to a classical diffusion process.
Jacob and Sach \cite{JS61}, in their field-theoretical analysis 
of a scalar particle coupled to two pions, found nonexponential terms 
decaying like $t^{-3/2}$ in the amplitude. Their explanation was 
geometrical: a particle produced at position $\vec{r}$ having velocity 
between $v$ and $dv$, will appear after a time $t$ within a 
spherical shell of radius $vt$ centered on $\vec{r}$, and thickness 
$tdv$. The probability that it will be found in 
a small volume element within the shell is inversely proportional 
to its volume $4\pi t^3v^2dv$. Hence the probability 
amplitude is proportional to $t^{-3/2}$. This  cannot be a universal 
explanation though, since the decay amplitude in 1D models 
behaves generically like $t^{-3/2}$, whereas the above argument 
translated to 1D would 
imply only $t^{-1/2}$. 

Post-exponential power-law behavior is sometimes interpreted as 
expressing the dominance of free-motion \cite{CSM77,Nuss}. However  
explicit calculations of the long-time propagator for specific 
potential scattering models show that in general both free and 
scattering terms are needed in the propagator, to reproduce the 
correct results \cite{MDS95j}. 

A general argument that justifies non-exponential decay is the 
so-called  ``initial state reconstruction'' \cite{Ersak69,FG72}.
Deviations from exponential decay would be the consequence of Feynman 
paths returning to the initial state from the orthogonal, decay 
products subspace. This argument alone, however, makes no 
quantitative predictions of the observed behavior and is applied to 
the survival probability, which is not always  easy to observe.

In this paper we take up 
and develop a frequently overlooked 
observation by R. G. Newton making use of classical mechanics 
\cite{Newton61,Newtonbook}. Newton noted that if a  
point source emits classical particles with an exponential decay law 
and with a suitable 
velocity distribution, the current density away from the source  
will eventually depend on time according to an inverse power law. 
Indeed, we 
shall show that by adjusting the parameters according to the 
quantum system, the classical model accurately reproduces 
the onset, power law, and intensity of post-exponential decay of the 
quantum probability density of a particle escaping from a trap. For 
simplicity we assume that the particle is restricted to the 
half-line, $r \ge 0$, as in quantum s-wave scattering. 
We shall also assume that the initial quantum state is orthogonal to any 
bound states so that it must eventually decay (escape) fully from the 
trap.   

%
%
%
%
%
\section{Classical and quantum sources and decay}
Consider first a source at $r=0$ which emits
classical particles with a definite velocity $v$ from $t_0=0$,
so that the fraction of particles 
emitted between $t_0$ and $t_0+dt_0$ is 
\beq                    
P(t_{0})dt_0=\frac{dt_0}{\tau}e^{-t_{0}/\tau},
\label{equ1}
\eeq
where $\tau$ is the emission life time. The spatial probability density 
observed at point $r$, at time $t$, is 
\beq                    
P_{c,v}(r,t)=\frac{1}{\tau v}e^{-(t-r/v)/\tau}\theta 
\left(t-\frac{r}{v}\right),
\eeq
where $\theta$ is the step function. For the more general case in which 
the emitted particles have a velocity distribution $\rho(v)$,  
%
%
\beq                    
P_{c}(r,t)=\int_{r/t}^\infty\!\!dv\,\rho(v)\frac{1}{v\tau}e^{-(t-r/v)/\tau},
\eeq
or, using $r=v(t-t_{0})$,
\beq                    
\label{e1}
P_{c}(r,t)=\int_{0}^t\!\!dt_{0}\, \frac{1}{(t-t_{0})\tau}\rho\!
\left(\frac{r}{t-t_{0}}\right)e^{-t_{0}/\tau}.
\eeq
The behavior for  $t>>>\tau$ of this integral can be expressed 
as an asymptotic series, 
\beq                    
P_{c}(r,t)\sim\sum_{n=0}^m\tau^{n}[g^{(n)}(0)-g^{(n)}(t)e^{-t/\tau}],
\eeq
where 
\beq                    
g(t_{0}) \equiv \frac{1}{t-t_0}\rho\!\left(\frac{r}{t-t_{0}}\right), 
\eeq
and $g^{(n)}$ is its $n$-th derivative with respect to $t_0$. 
The leading term asymptotically is 
\beq
\label{e2}              
P_{c}(r,t)\sim g(0)=\frac{1}{t}\rho\!\left(\frac{r}{t}\right).
\eeq
This is equivalent to Newton's result \cite{Newton61,Newtonbook}
(we use the probability
density rather than the current density).   
To advance from here, consider now the decay from a
quantum trap of a system prepared in 
a normalized non-stationary state $|\Psi_{0}\ra$. The wave function of 
this state at a point $r$ and time $t$ is 
\beq                    
\Psi(r,t)=\la r|e^{-iHt/\hbar}|\Psi_{0}\ra,
\eeq
with corresponding probability density $P_{q}(r,t)=|\Psi(r,t)|^{2}$. 
Using stationary states normalized in energy $u_{E}(r)$ (such that 
$\la u_{E^{'}}|u_{E}\ra=\delta(E-E^{'})$), and inserting the completeness 
relation,  
\beq                    
\Psi(r,t)=\int_{0}^{\infty}\!\!dE\,\la r|u_{E}\ra \la u_E|\Psi_0\ra  
e^{-iEt/\hbar}. 
\label{e10}
\eeq
%
%
%
The $u_E$ are solutions of the $s$-wave, radial Schr\"odinger equation,
\beq                    
\left[\frac{d^2}{dr^2}-v(r)+k^2\right]u_{E}(r)=0,
\eeq
where $v(r)=(2m/\hbar^2)V(r)$ and $k^2=(2m/\hbar^2)E$. As in 
\cite{MMS08}, it is convenient to define new solutions 
$w_{k}(r)=\hbar\sqrt{\frac{k}{m}}u_{E}(r)$ normalized as $\la 
w_{k^{'}}|w_{k}\ra=\delta(k-k^{'})$, which obey the boundary 
condition 
\beq
\label{e4}              
\lim\limits_{r\rightarrow\infty}w_{k}(r) = 
\sqrt{\frac{2}{\pi}}\sin [kr+\delta(k) ], 
\eeq
where $\delta(k)$ is the phase shift of the $s$-partial wave. These 
solutions are related to the regular solutions $\hat{\phi}_{k}(r)$ 
(which behave like the Riccati-Bessel function $\hat{j}_0(kr)$ as $r\to 0$), 
\beq
\label{e5}              
w_{k}(r)=\sqrt{\frac{2}{\pi}}\frac{\hat{\phi}_{k}(r)}{|f(k)|},
\eeq
where $f(k)= |f(k)| \exp (-i\delta)$ is the Jost function, as 
defined for example in Taylor \cite{Taylor}. It gives the relative 
normalization between solutions having unit incoming flux at 
infinity, and solutions that have slope $k$ at the origin. The 
partial-wave $S$-matrix element is $S(k)=f(-k)/f(k)$. Zeroes of 
$f(k)$ in the upper half complex momentum plane correspond to bound 
states, while those in the lower half plane are associated with 
scattering resonances. 

For an initially localized non-stationary state
\beqa                   
\la r|\Psi_{0}\ra=0 & \mbox{for} & r>r_{a},
\eeqa
the wave function can be written as
\beq
\label{e6}              
\Psi(r,t)=\frac{2m}{\pi\hbar^2}\int_{0}^{\infty}\!\!dE\, 
\frac{1}{k}\hat{\phi}_{k}(r)\frac{\la\hat{\phi}_{k}|\Psi_0\ra} 
{|f(k)|^2}e^{-iEt/\hbar}, 
\eeq
which has a form similar to the survival amplitude obtained in 
\cite{MMS08}. They have generically the same asymptotic behavior 
at long times, which corresponds to an energy distribution $\rho(E) = 
|\la u_E|\Psi_0\ra|^2 \sim E^{1/2}$, as $E \to 0$. This long time  
asymptotic behaviour is governed by the properties when $k \to 0$ of 
the integrand of Eq. (\ref{e6}).
For ``well behaved'' potentials, 
(those falling off faster than $r^{-3}$ when $r \to \infty$ and less 
singular than $r^{-3/2}$ at the origin), the $\ell = 0$ Jost function 
tends to a constant when $k \to 0$. (In the exceptional case that 
a zero energy resonance occurs $f(k=0)=0$.) The ${\hat \phi}_k$ 
behave near the origin as Ricatti-Bessel functions, ${\hat j_0(kr)}$, 
and are therefore linear in $k$. The behaviour of the integrand near 
threshold is thus $\sim E^{1/2}$. Following the same steps as in the 
derivation of the asymptotic behavior of the survival amplitude in 
\cite{MMS08}, one finds that the position probability density behaves 
like $P_q(r,t) \sim t^{-3}$ at long times. 


The energy distribution corresponds asymptotically to a velocity 
distribution since all particles are eventually released. The two 
distributions are related by  
\beq                    
\varrho(E)dE=\rho(v)dv.   
\eeq
Setting $E=\frac{1}{2}mv^{2}$, $m$ being the mass of the emitted particles, makes 
$\varrho(E)\sim E^{1/2} \Rightarrow  \rho(v)\sim v^{2}$.
Going back to Eq. (\ref{e2}) and considering the long time regime 
$t>>>\tau$, the classical particle velocity can be approximated by 
$v=r/t$, so $\rho(r/t)=\rho(v)$, which implies, as expected,  that 
at large $t$ 
the main contribution to the position probability density is from slow particles.
If we consider the same 
dependence as in the quantum case, $\rho(v)\sim v^{2}$, Eq. 
(\ref{e2}) implies an asymptotic behavior $P_{c}(r,t)\sim t^{-3}$, {\it i.e.},  
the classical model leads to the same  power law dependence as the 
quantum one. Moreover, in the following we shall see that it can be 
adjusted to provide the correct amplitude factor as well.    

Let us return to Eq. (\ref{e6}) and write the bra-ket factor as 
\beq                   
\label{e6b}
\la\hat{\phi}_{k}|\Psi_0\ra = \sqrt{\frac{\pi}{2}}|f(k)| 
\la  w_k|\Psi_0\ra,  
\eeq
%
Using Eq. (\ref{e5}) and the asymptotic behavior given 
by Eq. (\ref{e4}), the wave function for $r\rightarrow\infty$ 
may be written as
\beqa                   
\Psi(r,t) & \sim & 
\sqrt{\frac{2}{\pi}}\frac{m}{\hbar^{2}}\int_{0}^{\infty}\!\!dE\,\frac{1}{k}\ 
\la w_k|\Psi_0\ra \nonumber \\ 
\ & \ & \quad  \times \,\,\sin [kr+\delta(k)] e^{-iEt/\hbar}.
\eeqa
At low energy, the phase shift $\delta(k)$ is well described by 
the effective range expansion
\beq 
\label{era}            
k \cot \delta(k) = - \frac{1}{a_0} + \frac{1}{2} r_0 k^2 + \cdots, 
\eeq
where $a_0$ is the scattering length
while $r_0$ is called the effective 
range of the potential function.  
%
%
The asymptotic form of the resulting integral can be obtained from 
the Riemann-Lebesgue lemma \cite{R-L}. Only the main term which 
depends on the $k\to 0$ behavior of the integrand is kept. This 
gives a probability density of the form  
\beq                   
\label{e8}
P_{q}(r,t)\sim\beta\,[r - a_0]^{2}\frac{1}{t^{3}},
\eeq
where 
$\beta$ is 
the strength factor for the asymptotic dependence of the velocity 
distribution, $\rho(v)\sim \beta v^2$, which will depend on the 
particular state and potential. It has units of $[\beta] \sim v^{-3}$. In the 
approximation $v=r/t$ we can write 
\beq                   
\rho(v)\sim \beta\frac{r^{2}}{t^{2}}.  
\eeq
Introducing this in Eq. (\ref{e2}) we obtain for the classical 
probability density at long times 
\beq                  
\label{e9}
P_{c}(r,t)\sim\beta\frac{r^2}{t^{3}}.
\eeq
Compare now Eqs. (\ref{e8}) and (\ref{e9}), and to avoid confusion, 
let us rewrite $r\to r_{q}$ for the quantum case and $r\to r_{c}$ for 
the classical one. We see that if the classical coordinate is shifted 
by $a_0$, 
%
$r_{c}=r_{q} - a_0$, 
%
the classical model will reproduce the quantum probability density. 
Equivalently, $P_c(r)=P_q(r)$ at long times if the classical 
source is not at 
the origin but displaced by the scattering length $a_0$. 


In the exceptional case of a potential with a zero energy resonance $a_0
\to \infty$, and therefore the first term on the r.h.s. of Eq. (\ref{era})
is absent. This causes the Jost function to have a simple zero at $k=0$, see \cite{Taylor}, and therefore $|\la w_{k=0}|\Psi_0\ra|^2 $ is nonvanishing.
We then find, instead of Eq. (\ref{e8}), that 
\beq
\label{excep}
P_q(r,t)\sim |\la w_{k=0}|\Psi_0\ra|^2 m /(\hbar t),
\eeq
which is again in agreement with the classical expression (\ref{e2}) 
taking $\rho(v=0)=|\la w_{k=0}|\Psi_0\ra|^2 m /\hbar$.  
\section{Model calculation}
\begin{figure}[t]
\begin{center}
\includegraphics[height=3.5cm,angle=0]{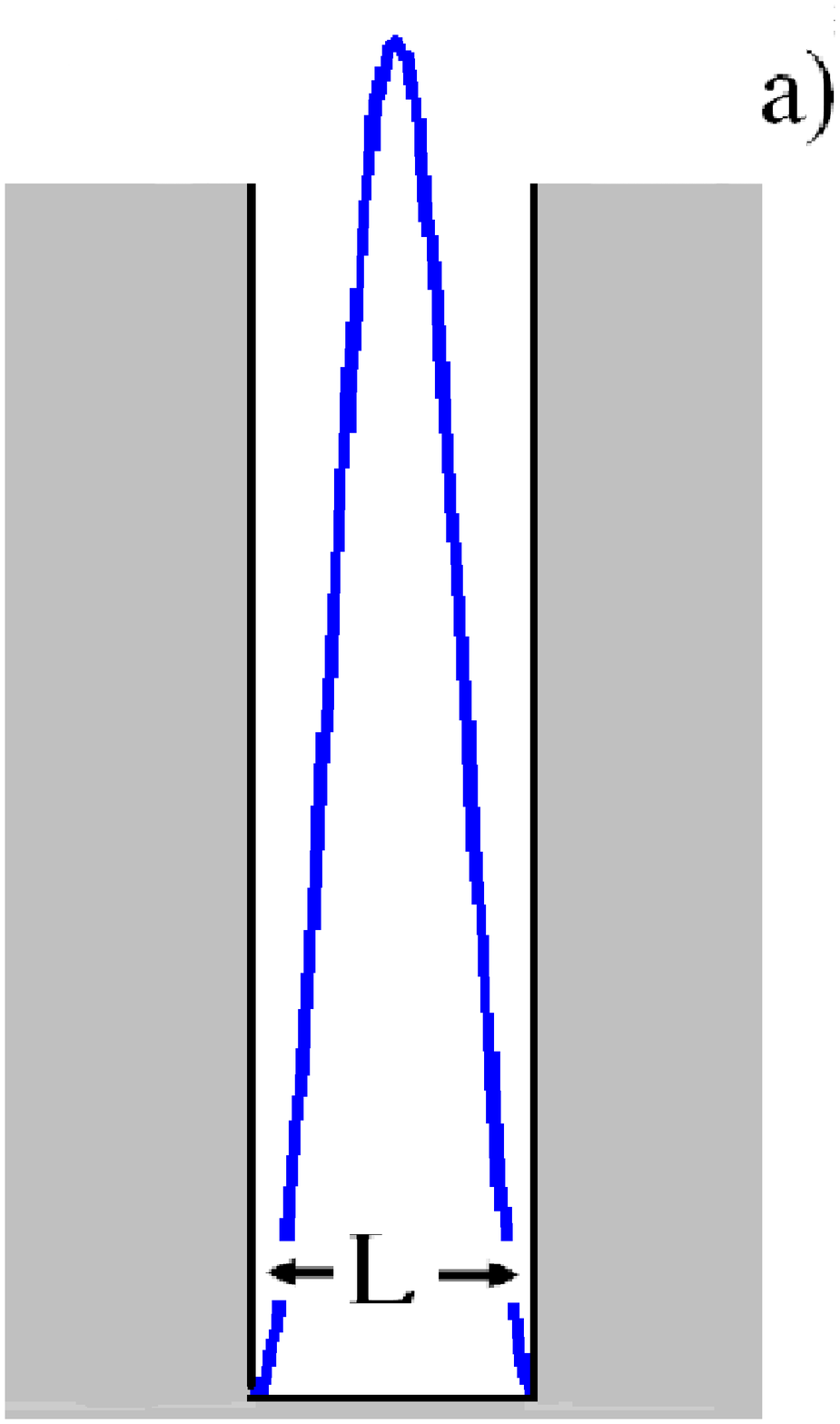}\hspace{0.25cm}
\includegraphics[height=3.35cm,angle=0]{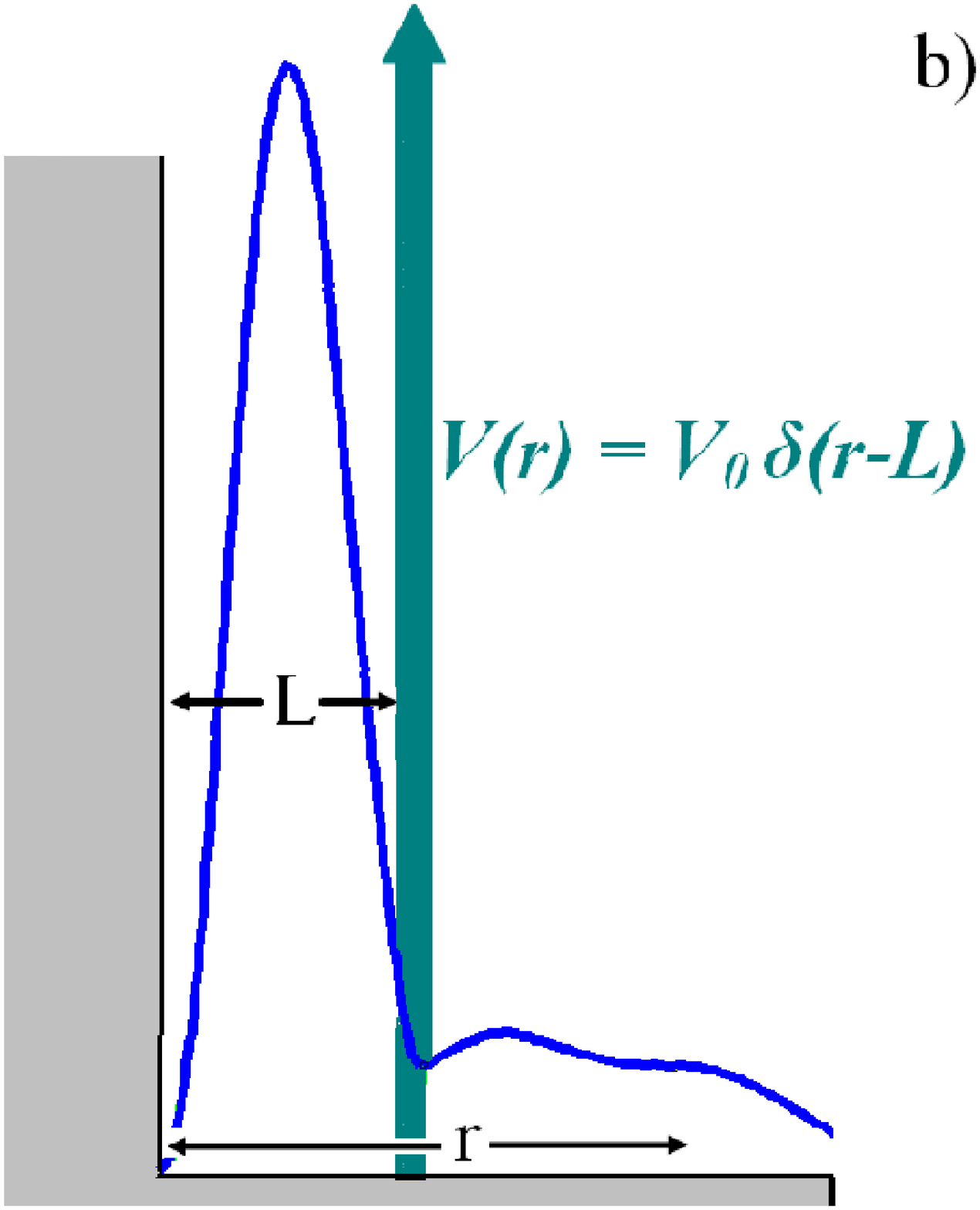}
\end{center}
\caption{\label{f2} 
(Color online) Scheme of the Winter model. a) The initial state is 
the ground state of an infinite square well. b) One of the 
walls is replaced by a delta barrier $V=V_{0}\delta(r-L)$.} 
\end{figure}

Now we check the above general results for Winter's 
delta-barrier model \cite{Winter} 
which is described in Fig. \ref{f2}.
%
%
The initial state is an eigenstate of the infinite square well potential,
\beq                    
\la r|\Psi_0\ra=\left\{
\begin{array}{cl}
\sqrt{\frac{2}{L}}\sin\left(\frac{n\pi r}{L}\right) & r\leq L 
\\
0 & r\geq L
\end{array}\right.,
\eeq
and the $k$-normalized basis functions are 
\beq                    
\la r|w_k\ra=e^{-i\delta(k)} \, 
\sqrt{\frac{2}{\pi}} 
\, \left\{
\begin{array}{lr} \sin(kr)/f(k) & r\leq L 
\\
(i/2)[e^{-ikr}-S(k)e^{ikr}] & r\geq L
\end{array}\right.,
\label{wf} 
\eeq
where $S(k)=f(k)^*/f(k)$, and the Jost function for this model is
\beq                    
f(k)=1+\frac{\alpha}{2ik}(e^{2ikL}-1),
\eeq
with  $\alpha=2mV_{0}/\hbar^{2}$.
%
%
$\rho(v)=|\la w_k|\Psi_0\ra|^{2}m/\hbar$ can be calculated exactly and takes the form
\beqa                   
\label{e11}
\rho(v)& = & \frac{Lm}{\pi\hbar}\frac{k^{2}}{k^{2}+\alpha k 
\sin (2kL) +\alpha^{2} \sin^{2}(kL)} \nonumber \\ 
&& \qquad \times \quad \left[\frac{2n\pi \sin(kL)}{k^{2}L^{2} - n^{2}\pi^{2}}\right]^{2}. 
\eeqa
{}From here 
the exact classical probability density is calculated numerically 
using Eq. (\ref{e1}), whereas the quantum density is given by the 
square modulus of Eq. (\ref{e10}). In the large-$t$ region the 
probability density has analytical expressions in both quantum and 
classical cases given by Eqs. (\ref{e8}) and (\ref{e9}) respectively. 
The coefficient $\beta$ is easy to find from Eq. (\ref{e11}) in the 
limit $v\to 0$, 
\beq                    
\beta=\frac{4m^3L^{3}}{(1+\alpha L)^{2}n^{2}\pi^{3}\hbar^3}.
\eeq
Also, Eq. (\ref{wf}) and 
$S(k)=e^{2i\delta(k)}$, give for the Winter model the explicit source 
shift 
\beq                    
\label{e12} 
a_0 = \frac{\alpha L^2}{1+\alpha L}, 
\eeq 
which, for $\alpha\ge 0$,  
lies between $0$ (for $\alpha L\to 0$, no barrier), and $L$ (for 
large $\alpha L$, strong confinement).

\begin{figure}
\begin{center}
\includegraphics[height=5cm,angle=0]{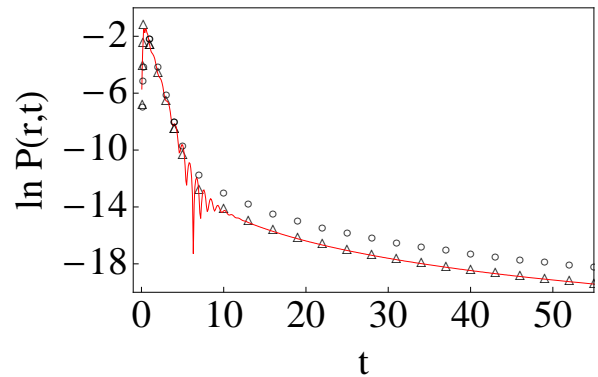}
\end{center}
\caption{\label{f3}
(Color online) The exact probability density obtained numerically 
versus $t$:  quantum result solid (red) line;  classical solution 
including the source shift, Eq. (\ref{e12})(triangles); classical 
solution without applying the source shift (circles). 
Parameter values: $\hbar=1$, $m=1/2$, $L=1$, $\alpha=5$, $n=1$, $r=2$, 
$\tau=0.5$.} 
\end{figure}

Finally, the quantum and classical probability densities 
both take (shifting the classical point source by $a_0$)
the post-exponential form
\beq                    
\label{e13}
P_{q,c}(r,t)\sim\frac{4}{n^{2}(1+\alpha L)^{2}}
\left(\frac{Lm}{\pi\hbar}\right)^{3}\left(r-
\frac{\alpha L^2}{1+\alpha L}\right)^2\frac{1}{t^{3}}. 
\eeq
%
%
%
The agreement is illustrated in Fig. \ref{f3},  where the exact 
decay curves (numerically integrated) are plotted. The classical 
density (triangles) indeed reproduces the quantum behavior (solid 
line) if the source shift is taken into account. For comparison we 
also show a curve in which the shift is not applied, so that the 
classical source remains at $r=0$ (circles). 
Taking the same value for $\tau$, 
we see that the classical model also agrees with the quantum one in 
the pre-exponential and exponential 
zones ($0 < t < 5$ in the drawing). The  classical model 
differs only in the absence of oscillations which occur 
at the onset of post-exponential behavior, 
due to quantum interference. 
The asymptotic behavior is indistinguishable on the scale of the 
figure from the analytical expression Eq. (\ref{e13}).  
\begin{figure}
\begin{center}
\includegraphics[height=5.5cm,angle=0]{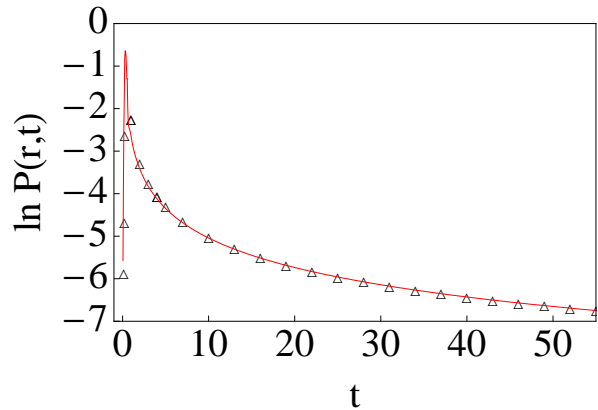}
\end{center}
\caption{\label{fze} 
(Color online) Logarithm of the probability density versus $t$, for an attractive delta potential 
having a zero energy resonance: exact quantum numerical solution (solid red line); classical model solution (triangles). 
The long-time behaviour is indistinguishable from Eq. (\ref{asy0er}).  
$\alpha=-1$, $\tau=0.2$, and other parameters as in Fig. 2.}
\end{figure}
\begin{figure}
\begin{center}
\includegraphics[height=5.5cm,angle=0]{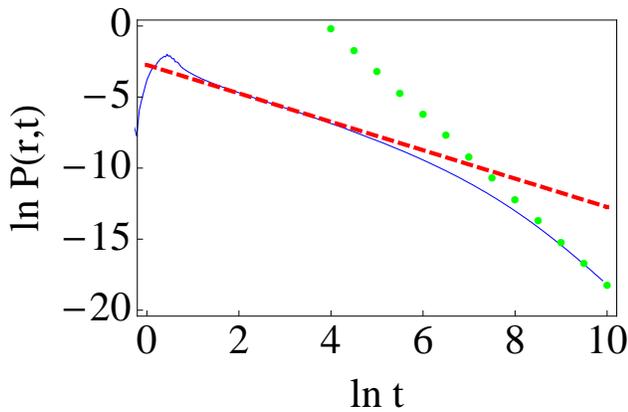}
\end{center}
\caption{\label{tra} 
(Color online) Logarithm of the probability density versus $t$ for a
delta potential slightly less attractive than required to produce 
a zero energy resonance:    
exact quantum numerical solution (solid blue line), and approximate power law 
decays proportional to $t^{-1}$ (Eq. (\ref{asy0er}), dashed red line) and $t^{-3}$ (Eq. (\ref{e13}), dotted green line).  
$\alpha=-0.98$, $r=10$, $\tau=0.2$, and other parameters as in Fig. 2.}
\end{figure}



%

The exceptional case of a zero energy resonance corresponds to an attractive 
delta with $\alpha=-1/L$. In this case from Eqs. (\ref{e2},\ref{excep}) we get  
\beq			
\label{asy0er}
P_{q,c}(x,t)\sim \frac{4 L m}{\hbar n^2 \pi^3 t}
\eeq
for both the classical and quantum cases: see Fig. \ref{fze}. 
If $\alpha$ is close to the critical value, say $\alpha=-1/L+\epsilon$, 
the decay follows a $t^{-1}$ decay law for some substantial period of time 
until the $t^{-3}$ decay eventually dominates, see Fig. \ref{tra}. 
The smaller is $\epsilon$, the longer does the $t^{-1}$ behaviour persist. 
\section{Discussion}
To summarize, the above results provide an intuitive physical
picture and quantitative description of post-exponential decay 
of the probability density at points distant 
from the source. We have developed 
the classical model suggested by 
Newton so as to achieve an accurate match between classical and 
quantum decays. Purely exponential decay from a source leads naturally, 
because of dispersion associated with a velocity distribution of 
the emitted particles, to the same power law decay in quantum and 
classical scenarios. Quantum mechanics is required 
to provide the emission characteristics, but initial state reconstruction (ISR) plays 
no role in the classical, purely outgoing dynamics. We have checked 
with the methodology of \cite{ISR}, that ISR-terms 
are negligible in the post-exponential range of times,  
in the quantum calculation of Fig. 2. This contrasts with their 
relevance to the survival probability \cite{ISR} and indicates   
different mechanisms for the transition to post-exponential decay 
inside and outside the source. Indeed, the survival probability 
(calculated either as 
$|\la\Psi(0)|\Psi(t)\ra|^2$ or $|\la r|\Psi(t)\ra|^2$  with $r< L$ )
is still in its exponential regime when the transition shown in Fig. 2 
(at $r=2$)  
takes place, {\it i.e.}, the purely exponential decay hypothesis (\ref{equ1}) 
for the classical source 
is justified, and the onset of the post-exponential regime of survival 
within the trap cannot causally 
affect the transition observed in the density outside the source.


Due to recent advances in lasers, semiconductors, nanoscience, and 
cold atoms, microscopic interactions are now relatively easy to 
manipulate, decay parameters have become controllable, and 
post-exponential decay more accessible to experimental scrutiny 
and/or applications \cite{TMMS}. Under appropriate conditions it 
could become the dominant regime and be used to speed-up decay 
{\it via} an Anti-Zeno effect \cite{Lewenstein}. Moreover, recent 
experiments on periodic waveguide arrays provide a classical, 
electric field analog of a quantum system with exponential decay 
\cite{Lon06a,VLL07}, where the post-exponential region could be 
studied in a particularly direct way. 

\section*{Acknowledgments}
J. G. M. acknowledges the kind hospitality of the Max Planck 
Institute for Complex Systems in Dresden. We acknowledge 
funding by the Basque Country University UPV-EHU (GIU07/40), the  
Ministerio de Educaci\'on y Ciencia Spain (FIS2006-10268-C03-01/02, and FIS2009-12773-C02-01), 
and NSERC Canada (RGPIN-3198). 
E. T. acknowledges financial support by the Basque Government (BFI08.151).


\begin{thebibliography}{99}  
\bibitem{KD08} L. M. Krauss and J. Dent, Phys. Rev. Lett. {\bf 100}, 171301
(2008).

\bibitem{Winterh} R. G. Winter,  Phys. Rev. {\bf 126}, 1152 (1962).

\bibitem{NB77} C. A. Nicolaides and  D. R. Beck, Phys. Rev. Lett. {\bf 38}, 683, 1037
(1977).

\bibitem{Nicolaides02} C. A. Nicolaides, Phys. Rev. A {\bf 66}, 022118 (2002). 

\bibitem{No95} E. B. Norman, B. Sur, K. T. Lesko, R. M. Larimer, D. J. DePaolo,   
and T. L. Owens, Phys. Lett. B {\bf 357}, 521 (1995).

\bibitem{RHM06} C. Rothe, S. I. Hintschich,  and A. P. Monkman, Phys. 
Rev. Lett. {\bf 96}, 163601 (2006).

\bibitem{book} J. Martorell, J. G. Muga, and D. W. L. Sprung, in 
{\it Time in Quantum Mechanics, vol. 2}, ed. by J. G. Muga, A. Ruschhaupt, and A. del Campo (Springer, Berlin, 2009).

\bibitem{TMMS} E. Torrontegui, J. G. Muga, J. Martorell and D. W. L. Sprung, 
Phys. Rev. A {\bf 80}, 012703 (2009).

\bibitem{Newton61} R. G. Newton, Ann. Phys. (NY) {\bf 14}, 333 (1961).

\bibitem{Newtonbook} R. G. Newton, Scattering Theory of Waves and Particles, 
Second Ed. (Dover, Mineola, 2002). 

\bibitem{Khalfin} L. A. Khalfin, Soviet Physics JETP {\bf 6}, 1053 
(1958). 

\bibitem{Petzold} J. Petzold, Z. Phys. {\textbf{155}}, 422 (1959). 

\bibitem{Winter} R. G. Winter,  Phys. Rev. {\bf 123}, 1503 (1961).

\bibitem{Lan} R. Landauer, private conversation with J. G. Muga.  

\bibitem{Hellund53} E. J. Hellund, Phys. Rev. {\bf 89}, 919 (1953).

\bibitem{JS61} R. Jacob  R. G. Sachs, Phys. Rev. {\bf 121}, 350 
(1961). 

\bibitem{CSM77} C. B. Chiu, E. C. G. Sudarshan, B. Misra, Phys. Rev. 
D {\bf 17}, 520 (1977).

\bibitem{Nuss} H. M. Nussenzveig,  {\it Causality and Dispersion Relations} 
(Academic Press, New York, 1972).

\bibitem{MDS95j} J. G. Muga, V. Delgado,  R. F. Snider, 
Phys. Rev. B {\bf 52}, 16381 (1995). 

\bibitem{Ersak69} L. Ersak, Yad. Fiz. {\bf 9}, 458 (1969); English translation: Sov. J. Nucl. Phys. {\bf 9}, 263 (1969). 

\bibitem{FG72} L. Fonda,  G. C. Ghirardi, Il Nuovo Cimento {\bf 7A}, 180 (1972)

\bibitem{MMS08} J. Martorell, J. G. Muga, and  D. W. L. Sprung, Phys. Rev. A {\bf 77}, 042719 (2008).

\bibitem{Taylor} J. R. Taylor, {\it Scattering Theory} (Wiley, New York, 1972).

\bibitem{R-L} E. T. Whittaker and G. N. Watson, {\it A Course of Modern Analysis} (Cambridge, fourth edn. 1927) p. 172.

%
\bibitem{ISR} J. G. Muga, F. Delgado, A. del Campo,
and G. Garc\'\i a-Calder\'on, Phys. Rev. A {\bf 73}, 052112 (2006).

\bibitem{Lewenstein} M. Lewenstein and  K. Rzazewski, Phys. Rev. A {\bf 61}, 022105
(2000). 

\bibitem{Lon06a} S. Longhi, 
Phys. Rev. Lett. {\bf 97}, 110402 (2006).

\bibitem {VLL07} G. Della Valle, S. Longhi, P. Laporta, P. Biagioni,
L. Dou, and M. Finazzi, 
Appl. Phys. Lett. {\bf 90}, 261118 (2007).
%
\end{thebibliography}
\end{document}